\documentclass[]{aastex62}
\usepackage{amsmath}
\usepackage{appendix}
\usepackage{multirow}
\usepackage{graphicx}
\usepackage{subfigure}
\usepackage{threeparttable}

\newcommand{\kms}{${\rm km~s}^{-1}$}
\newcommand{\Usun}{$U_{\odot}$}
\newcommand{\Vsun}{$V_{\odot}$}
\newcommand{\Wsun}{$W_{\odot}$}
\begin{document}
\title{Kinematics of the Local Spiral Structure Revealed by Young Stars in \emph{Gaia}~DR3}
\author{Dejian Liu}, 
\affiliation{Purple Mountain Observatory, Chinese Academy of Sciences, Nanjing 210023, People's Republic of China}
\affiliation{School of Astronomy and Space Science, University of Science and Technology of China, Hefei 230026, People's Republic of China}
\author{Ye Xu}
\affiliation{Purple Mountain Observatory, Chinese Academy of Sciences, Nanjing 210023, People's Republic of China}
\affiliation{School of Astronomy and Space Science, University of Science and Technology of China, Hefei 230026, People's Republic of China}
\author{Chaojie Hao}
\affiliation{Purple Mountain Observatory, Chinese Academy of Sciences, Nanjing 210023, People's Republic of China}
\affiliation{School of Astronomy and Space Science, University of Science and Technology of China, Hefei 230026, People's Republic of China}
\author{Shuaibo Bian}
\affiliation{Purple Mountain Observatory, Chinese Academy of Sciences, Nanjing 210023, People's Republic of China}
\author{Zehao Lin}
\affiliation{Purple Mountain Observatory, Chinese Academy of Sciences, Nanjing 210023, People's Republic of China}
\affiliation{School of Astronomy and Space Science, University of Science and Technology of China, Hefei 230026, People's Republic of China}
\author{Yingjie Li}
\affiliation{Purple Mountain Observatory, Chinese Academy of Sciences, Nanjing 210023, People's Republic of China}
\author{Jingjing Li}
\affiliation{Purple Mountain Observatory, Chinese Academy of Sciences, Nanjing 210023, People's Republic of China}

\begin{abstract}
Using young open clusters and O--B2-type stars in~\emph{Gaia}~DR3, we investigate 
the kinematics of the local spiral structure. 
In general, the young sources in the outer spiral arms may present larger peculiar motions than those in 
the inner spiral arms.
The young open clusters appear to have smaller peculiar motions than the O--B2-type stars, and the sources 
in both the Perseus and Local Arms may show an inward motion toward the Galactic center 
and rotate slower than Galactic rotation. Meanwhile, the sources in the Carina Arm may move in the opposite direction from the Sun to the Galactic center and rotate marginally faster than Galactic rotation.
In addition, using young open clusters and O--B2-type stars, we have improved the distance 
estimations of kinematic methods for several regions near the Sun.
\end{abstract}

\section{Introduction}

The kinematics of the spiral arms of the Milky Way are of great interests to 
astronomers, as they can help unveil the formation and evolution of our home Galaxy.
As key elements of the spiral arms, the measurements and analysis of 
the spatial distributions, velocities, and rotational orbits of 
stars and gas therein can provide valuable insights into Galactic structure and 
dynamics~\citep{Binney+Tremaine+2008}, as well as the gravitational 
potential of the Milky Way~\citep{Sakai+etal+2019,Immer+etal+2019}.

Numerous young objects have been used to trace the spiral-arm structure of 
our Galaxy, e.g., 
high-mass star-formation region (HMSFR) masers yielded by the Bar and Spiral Structure 
Legacy (BeSSeL) Survey~\citep{brunthaler2011,Reid+etal+2014,Reid+etal+2019} and the Very Long Baseline Interferometry (VLBI) Exploration of 
Radio Astrometry (VERA) array~\citep{VERA+2020},
as well as observations of young massive O--B-type stars~\citep{Xu+etal+2018,Xu+etal+2021}, 
H\textsc{ii} regions~\citep{Georgelin+Georgelin+1976,Hou+Han+2014}, 
young open clusters~\citep[YOCs,][]{Dias+etal+2019,Hao+etal+2021,
GaiaMW+2022}, etc.
Although the global Galactic spiral structure is still a topic of debate, 
the spiral structure within about 5~kpc of the Sun is better understood.
There are three spiral segments in the solar neighborhood, namely the 
Sagittarius--Carina, Local, and Perseus Arms~\citep{Reid+etal+2019}. 
Recently, \citet{Xu+etal+2023} reconsidered the spiral structure of the Milky Way and 
proposed that the Sagittarius Arm may be not connected to the Carina Arm.
Due to a lack of available sources in the Sagittarius Arm, in this work, we 
mainly concentrate on the three distinct spiral-arm segments 
in the solar neighborhood: the Carina, Local, and Perseus Arms.

The kinematics of the Milky Way are closely related to its structure. In recent years, substantial progress in studying the local kinematics using stars provided by \emph{Gaia} and VLBI masers has been achieved. 
Using {\it Gaia} data release 2~\citep{gaia+2018}, \citet{GaiaMW+2018} looked at the kinematics of the Milky Way disk, within a radius of several kiloparsecs around the Sun, and found rich and complex kinematics of the Galactic disk and streaming motions in all three dimensional velocity components.
\citet{GaiaMW+2022} studied the Galactic spiral structure and kinematics using young stellar populations in \emph{Gaia} data release 3~\citep[DR3,][]{GaiaDR3+2022}, and revealed that the velocity field of the OB stars shows streaming motions whose characteristic length is similar to the spiral arm density.
As reviewed by~\citet{Immer+Rygl+2022}, based on VLBI masers, researchers have 
studied the kinematics of sources located in different spiral arm segments.
However, many of these studies focused only on the peculiar motion of masers in individual 
spiral arm segments, while systematic kinematic studies of the local spiral structure, 
especially using young sources located in the spiral arms, are still very lacking.
In this work, based on the spiral-arm model proposed by~\citet{Xu+etal+2023},
we aim to study the kinematics of the local spiral structure using young stars.

Accurate trigonometric parallaxes and proper motions of HMSFR masers in the 
northern hemisphere sky have been determined~\citep[e.g.,][]{Reid+etal+2019,Xu+etal+2021,bian2022}.
The high-precision six-dimensional (6D; 3D spatial and 3D velocity) 
astrometric parameters of OB stars and YOCs, provided by~\emph{Gaia}~DR3,
enable us to investigate the kinematics of young stars in 
the local spiral structure, and compare them with those of HMSFR masers.
Additionally, based on the astrometric parameters of young stars in the solar 
neighborhood, we try to provide valuable indications for future distance 
estimation using kinematic methods.

\section{Data}
\subsection{YOCs}

\citet{Xu+etal+2023} compiled a sample that contains more than 5\,000 open clusters (OCs)
from three catalogs~\citep{Hao+etal+2021,Hao+etal+2022a,Castro+etal+2022}. 
In this work, we increase this catalog by considering the 38 newly reported OCs
by~\citet{Hao+etal+2022b} and the 101 newly reported OCs by~\citet{Qin+etal+2023}. 
We also identify potentially duplicate OCs.
If the difference between the mean parameters of two OCs are within 
3$\sigma_{i}$ ($i = \alpha, \delta, \varpi, 
\mu_{\alpha} \cos \delta, \mu_{\delta}$) in five-dimensional parametric space,
they can be regarded as duplicate objects, and we confirm them via visual inspection.
After this step, 4\,871 OCs remain (see Table~\ref{tab:OCs}), of which 3\,794 have \emph{Gaia} radial velocity 
measurement. 
In the following analysis, we only use young OCs (age $<$ 20 Myr) with parallax accuracies 
better than 20\% and radial velocity uncertainties smaller than 10~\kms.
Determination of the radial velocity uncertainties of OCs is conducted following the same method
as~\citet{Soubiran+etal+2018}. 
Finally, we obtain a sample containing 285 YOCs, and their 
astrometric parameters are taken from \emph{Gaia}~DR3. We perform a 
parallax zero-point correction for the member stars of each OC~\citep{Lindegren+etal+2021},
and then use the average parallax of all members as the OC’s parallax.
\begin{deluxetable}{ccccccccccccccc}[htbp]
    \tablecaption{Summary of the Mean Parameters of the Cluster Sample
    \label{tab:OCs}
    }
    \tabletypesize{\tiny}
    \tablehead{
     & & & & & & & & & & & \\
    cluster & $N$ & $\alpha$ & $\delta$ & $\varpi$ & $\sigma_{\varpi}$ & $\mu_{\alpha} \cos \delta$ & $\sigma_{\mu_{\alpha} \cos \delta}$ & $\mu_{\delta}$ & $\sigma_{\mu_{\delta}}$ & $v_{\rm los}$ & $\sigma_{v_{\rm los}}$ & $N_{v_{\rm los}}$ & $\log t$ & Ref \\
    & & (deg) & (deg) & (mas) & (mas) & (mas yr$^{-1}$) & (mas yr$^{-1}$) & (mas yr$^{-1}$) & (mas yr$^{-1}$) & (\kms) & (\kms) & & & }     
    \startdata
    UBC1194 & 64 & 0.020  & 63.900 & 0.330 & 0.023 & -2.496 & 0.024 & -1.261 & 0.023 & -77.9 & 10.7 & 5 & 7.78 & Castro2022 \\
    Berkeley\_58 & 115 & 0.063 & 60.937 & 0.331 & 0.084 & -3.361 & 0.091 & -1.705 & 0.089 & -4.1 & 10.0 & 1 & 8.47 & Scholz2015 \\
    NGC\_7801 & 33 & 0.077 & 50.711 & 0.444 & 0.064 & -3.642 & 0.056 & -2.439 & 0.075 & -12.5 & 14.4 & 4 & 9.26 & Scholz2015 \\
    ...
    \enddata
    \tablecomments{This table is available in its entirety in machine-readable form.}
\end{deluxetable}

\subsection{O--B2-type Stars}

We adopt the method of~\citet{Xu+etal+2023} to identify O--B2-type stars 
in \emph{Gaia}~DR3 and a renormalized unit weight error (\texttt{ruwe})
$< 1.4$~\citep{Lindegren+etal+2021} is used to remove objects with unreliable 
astrometric solutions. 
After that, 1\,135 O--B2-type stars with \emph{Gaia} radial velocities are selected (see Table~\ref{tab:OB}).
The values of the parameter \texttt{ipd\_gof\_harmonic\_amplitude} above 0.1 in combination with \texttt{ruwe} being larger than 1.4 are indicative of resolved binaries, which are still not correctly handled in the astrometric processing, and may cause spurious solutions~\citep{gaia2021}. Hence, these two parameters are adopted to reduce the number of potential binaries, and then reduce the effect due to them, i.e., \texttt{ipd\_gof\_harmonic\_amplitude} $< 0.1$ and \texttt{ruwe}
$< 1.4$~\citep{Fabricius+etal+2021}.
Applying the same criteria of a parallax precision better than 20\% and a radial velocity error smaller than 10~\kms, we obtained a final sample of 300 O--B2-type stars.
Same as the OCs, we also perform a parallax zero-point correction for each O--B2-type star in the sample~\citep{Lindegren+etal+2021}.

\begin{deluxetable}{cccccccccccc}[htbp]
    \tablecaption{Summary of the Parameters of the O--B2-type Stars
    \label{tab:OB}
    }
    \tabletypesize{\tiny}
    \tablehead{
     & & & & & & & & & & & \\
    \emph{Gaia} DR3 ID & $\alpha$ & $\delta$ & $\varpi$ & $\sigma_{\varpi}$ & $\mu_{\alpha} \cos \delta$ & $\sigma_{\mu_{\alpha} \cos \delta}$ & $\mu_{\delta}$ & $\sigma_{\mu_{\delta}}$ & $v_{\rm los}$ & $\sigma_{v_{\rm los}}$ & $T_{\rm eff}$ \\
    & (deg) & (deg) & (mas) & (mas) & (mas yr$^{-1}$) & (mas yr$^{-1}$) & (mas yr$^{-1}$) & (mas yr$^{-1}$) & (\kms) & (\kms) & (K)}
    \startdata
    528594342521399168 & 0.4453 & 67.5070 & 1.010 & 0.012 & -1.566 & 0.011 & -1.795 & 0.012 & -42.7 & 19.3 & 42190\\
    429366957178085376 & 1.0123 & 60.8458 & 0.303 & 0.023 & -2.544 & 0.018 & -1.969 & 0.021 & -51.0 & 34.9 & 24480 \\
    431544127639952384 & 1.9834 & 63.3354 & 0.427 & 0.018 & -2.823 & 0.021 & -0.925 & 0.020 & -55.5 & 6.9 & 20000 \\
    ...
    \enddata
    \tablecomments{This table is available in its entirety in machine-readable form.}
\end{deluxetable}

\section{Kinematics of Young Objects}

Using the parallaxes to convert proper motions to linear velocities and combining them 
with the radial velocities, we derive the peculiar motions ($U_{s}, V_{s}, {\rm and}~W_{s}$) of the
YOCs and O--B2-type stars, using the same method as~\citet{Reid+etal+2009}. 
$U_{s}, V_{s}$, and $W_{s}$ are the velocity components towards the Galactic center (GC, radial), in the direction of Galactic rotation ( azimuthal), and towards the north Galactic pole, respectively. 
In the above calculation, the fundamental parameters of the Milky Way, $R_{0} = 8.15$~kpc, 
$\Theta_{0} = 236$~\kms, and solar motions of \Usun~= 10.6~\kms, 
\Vsun~= 10.7~\kms, and \Wsun~= 7.6~\kms~\citep{Reid+etal+2009} are adopted. 
Meanwhile, a universal rotation curve~\citep{Persic+etal+1996} consisting 
of two parameters, a2 = 0.96 and a3 = 1.62~\citep{Reid+etal+2019}, is assumed.

\begin{figure}[htbp]
    \centering
    \subfigure[YOCs]{\includegraphics[width = 8cm]{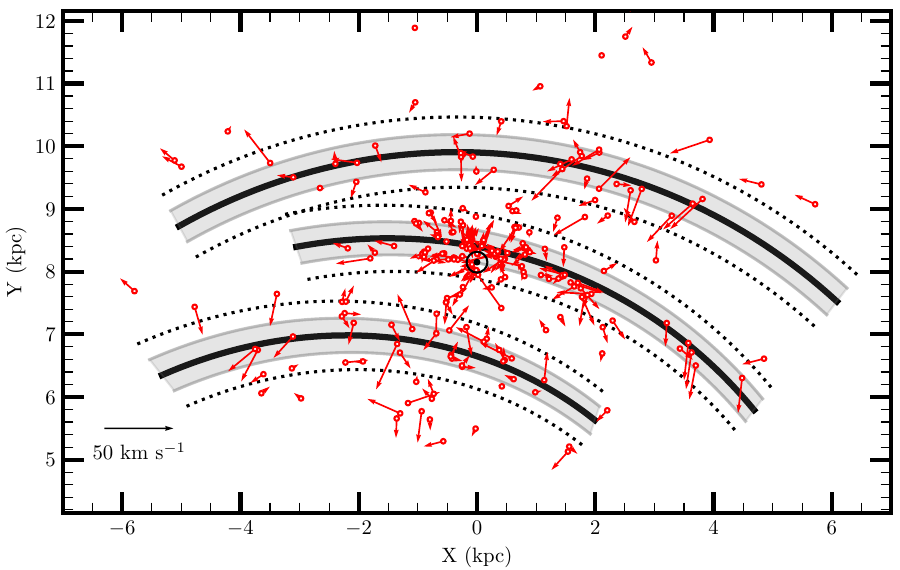}}
    \subfigure[O--B2-type stars]{\includegraphics[width = 8cm]{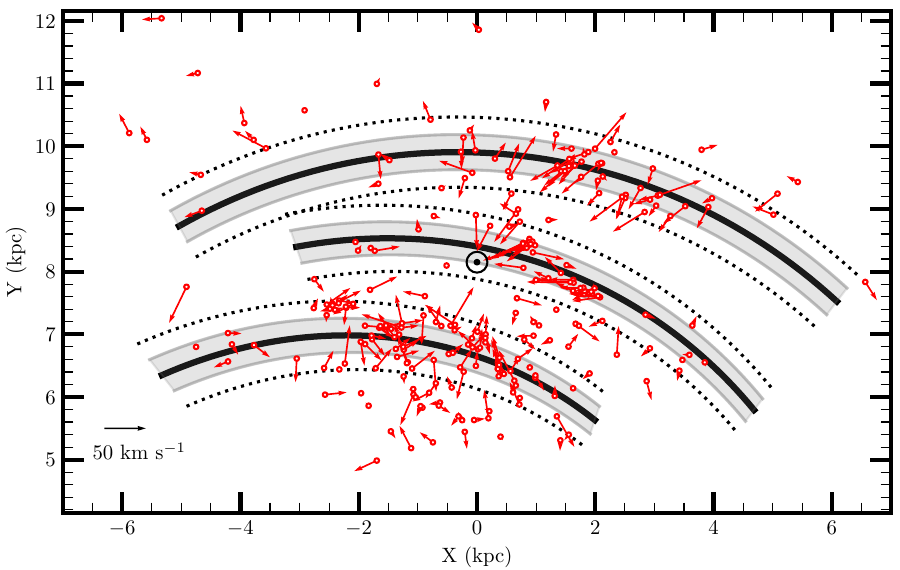}}
    \caption{Peculiar motions of the YOCs (left) and O--B2-type stars (right) in the Galactic plane. The parallax uncertainties are smaller than 20\% and the radial velocity uncertainties are smaller than 10~\kms. The solid curved lines trace the centers of the spiral arms: from top to bottom are the Perseus, Local, and Carina Arms, respectively~\citep{Xu+etal+2023}. The shadows and dashed lines are the $1 \sigma$ and $2 \sigma$ widths from the center of the spiral arms, respectively. The Sun (black Sun symbol) is at (0, 8.15)~kpc~\citep{Reid+etal+2019}. A 50~\kms~scale vector is shown in the lower-left corner. }
    \label{fig:OC_OB}
\end{figure}

Figure~\ref{fig:OC_OB} displays the peculiar motions of the YOCs and O--B2-type stars in 
the Galactic plane, which are almost concordant with each other.
There are two significant kinematic features of sources in the Perseus and Local Arms.
Near the direction of $l$ $\sim$ $100^{\circ}$--$135^{\circ}$,
the YOCs and O--B2-type stars in the Perseus Arm exhibit large peculiar motions,
which is consistent with the result shown using HMSFR masers~\citep{Reid+etal+2019}.
Around the direction of $l~\sim70^{\circ}$--$90^{\circ}$, both the YOCs and O--B2-type stars in the Local Arm exhibit opposite motions with respect to Galactic rotation, 
similar to the results of HMSFR masers~\citep{Xu+etal+2013}.
Therefore, these results suggest that the peculiar motions of the different young populations in the solar 
neighborhood probably are comparable. 

\begin{figure}[htbp]
    \centering
    \subfigure[O-type stars]{\includegraphics[width = 8cm]{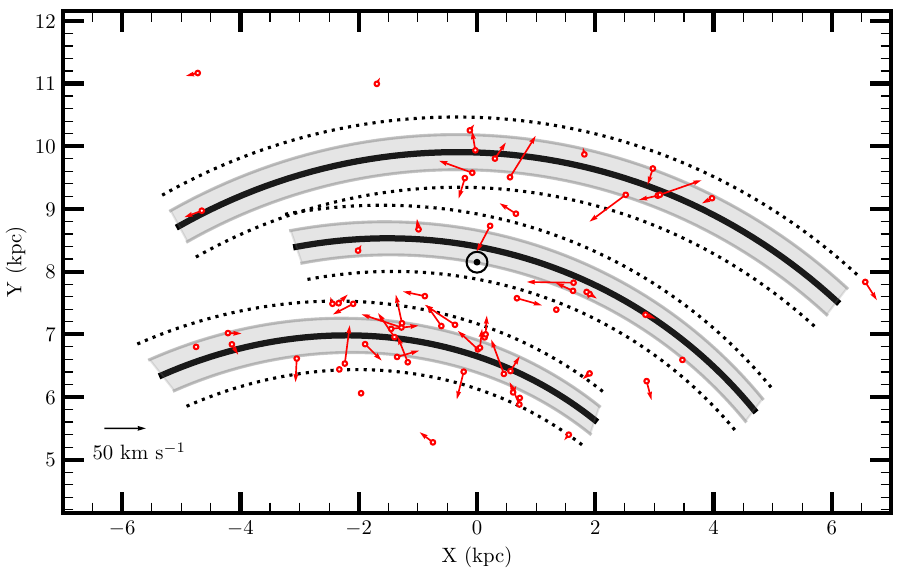}}
    \subfigure[B0--B2-type stars]{\includegraphics[width = 8cm]{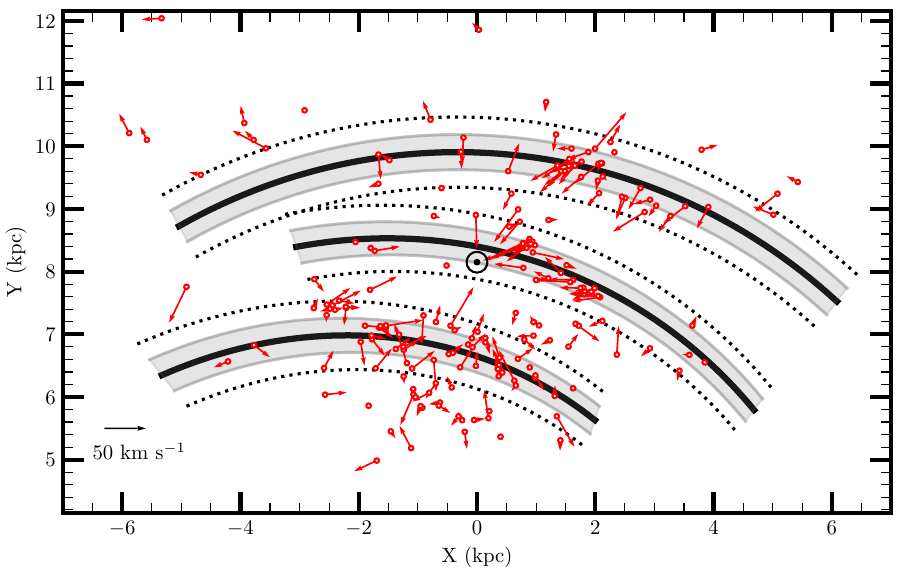}}
    \caption{Peculiar motions of the O (left) and B0--B2 (right) stars in Galactic plane. The others features are the same as in Figure~\ref{fig:OC_OB}.}
    \label{fig:O_B}
\end{figure}

To address the issue of whether there are kinematic differences between young stars in different 
evolutionary stages, we divide the O--B2-type stars into the O-type stars and B0--B2-type stars 
based on their effective temperature, where the O-type stars are hotter than 30\,000~K and 
B0--B2-type stars are hotter than 20\,000~K but colder than 30\,000~K~\citep{Chen+etal+2013}.
As shown in Figure~\ref{fig:O_B}, we present the peculiar motions of the O- and B0--B2-type 
stars in the Galactic plane. 
It appears that the B0--B2-type stars, overall, possess larger peculiar motions than those of the O-type stars.

Next, we investigate the kinematic characteristics of the YOCs and O--B2-type stars in 
each spiral arm near the Sun, and analyze the average peculiar motion ($\overline{U_{s}}$, $\overline{V_{s}}$, and $\overline{W_{s}}$) of those sources with respect to the 
Galactocentric radius $R$. 
Following~\citet{Reid+etal+2019}, we define an outlier as having greater than a 
3$\sigma$ residual in any peculiar motion 
component and, based on this, we remove 20 YOCs and 44 O--B2-type stars from the following analysis.

\subsection{Kinematics of Young Objects in the Spiral Arms}

\begin{figure}
    \centering
    \includegraphics[width = 5.5cm]{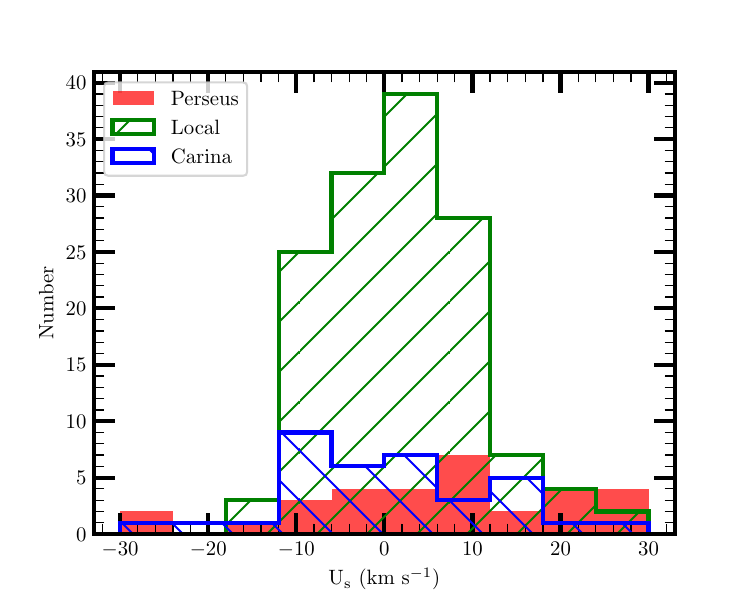}
    \includegraphics[width = 5.5cm]{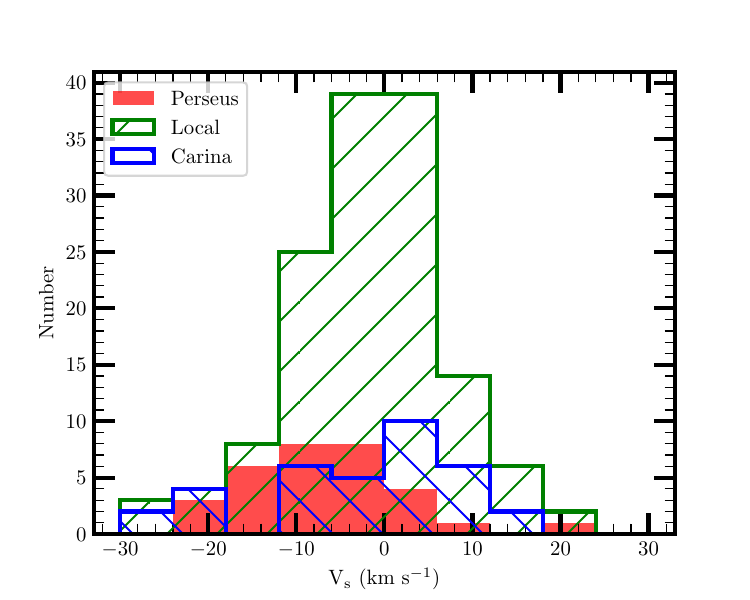}
    \includegraphics[width = 5.5cm]{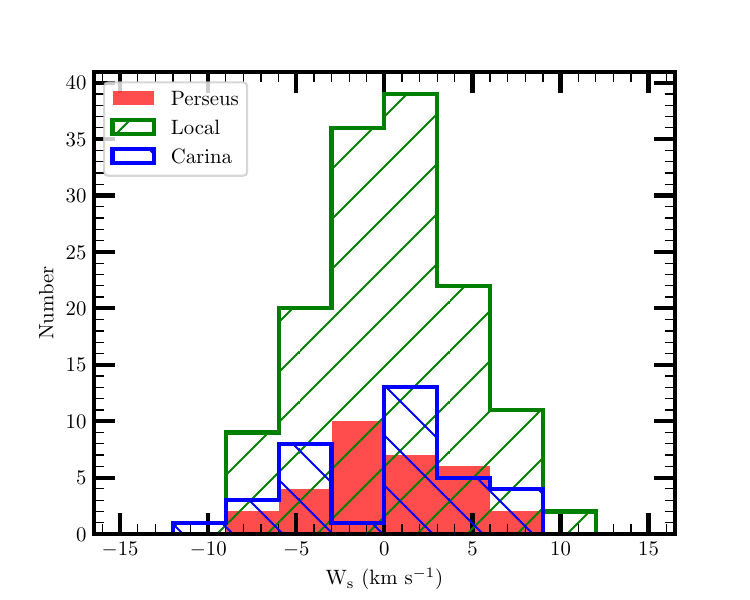}
    \caption{Histograms of peculiar motions of the YOCs.}
    \label{fig:UVW_hist_OC}
\end{figure}

\begin{figure}
    \centering
    \includegraphics[width = 5.5cm]{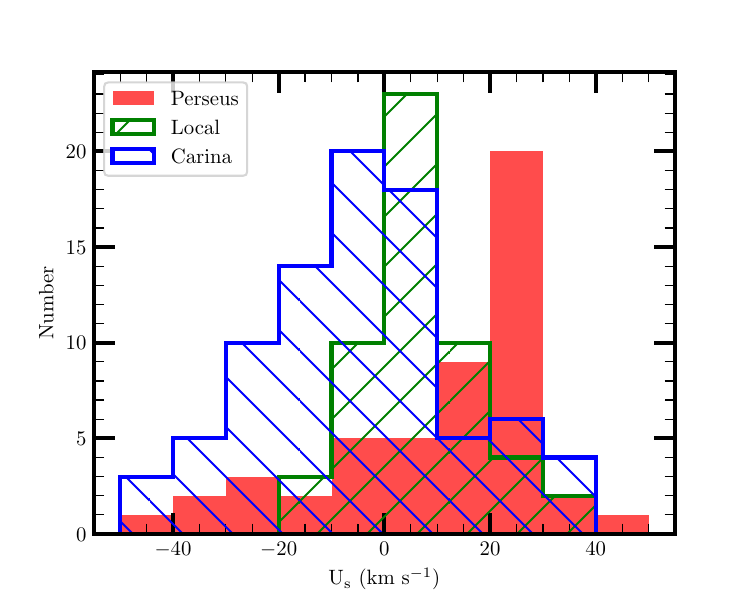}
    \includegraphics[width = 5.5cm]{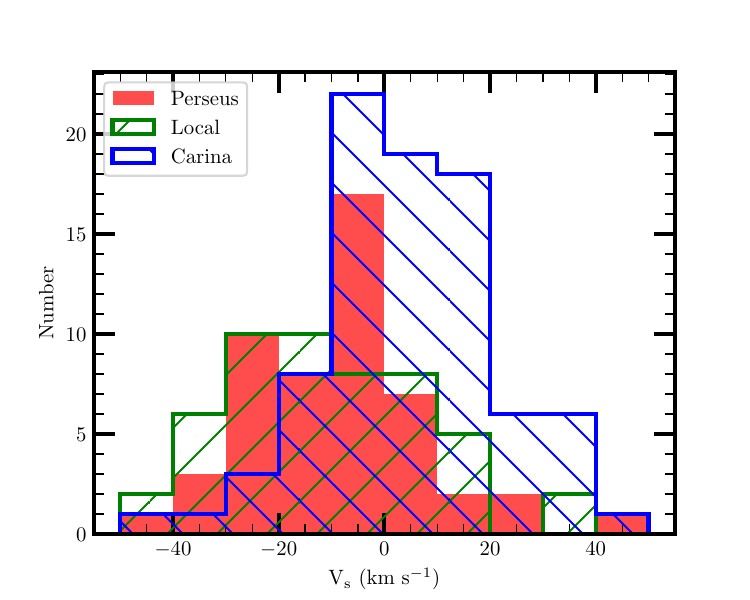}
    \includegraphics[width = 5.5cm]{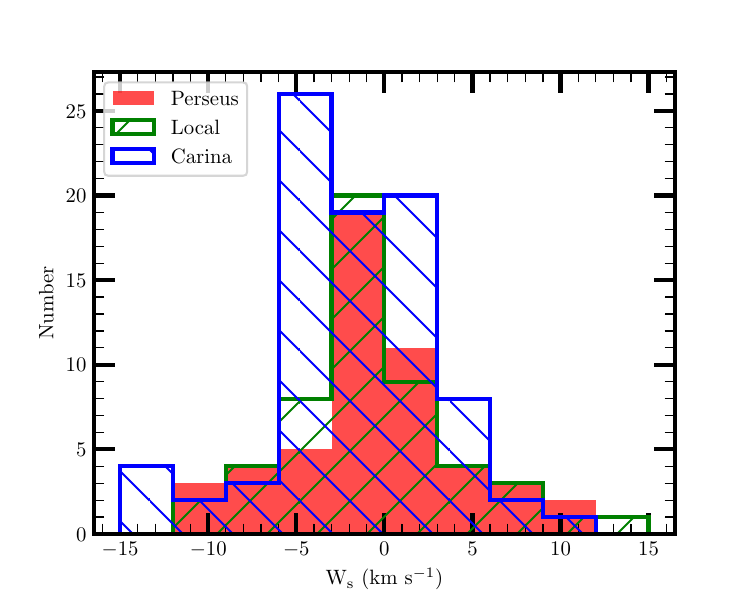}
    \caption{Histograms of peculiar motions of the O--B2 type stars.}
    \label{fig:UVW_hist_OB}
\end{figure}

In this work, we focus on analyzing the 
kinematic characteristics of sources in the three spiral arms near the Sun, including the Perseus, Local, and Carina Arms. 
Besides, by comparing the kinematics of sources at different distances (i.e., $1 \sigma$ and $2 \sigma$) from the center of each spiral arms, we can investigate the areas most influenced by the gravitational force of each spiral arms. Adopting the spiral-arm model proposed by~\citet{Xu+etal+2023}, the average peculiar motions and std values are similar for sources located within $1 \sigma$ and $2 \sigma$ widths from the center of each spiral arm, so we just use the $2 \sigma$ for this study.
In Figures~\ref{fig:UVW_hist_OC} and \ref{fig:UVW_hist_OB}, we show histograms of peculiar motions of YOCs and O--B2-type stars.
Then, we calculate the average peculiar motions and corresponding errors, and the standard deviation (std) of the peculiar motions of sources in each spiral arm, which are listed in Tables~\ref{tab:UVW_arms} and~\ref{tab:UVW_arms_std}. 
For the 199 HMSFR masers summarized by~\citet{Reid+etal+2019}, and the five newly reported by~\citet{Xu+etal+2021} and~\citet{bian2022}, \citet{Xu+etal+2023} re-assigned them to the spiral arms in their new proposed spiral-arm model. Here, the masers belonging to the Perseus Arm in the second and third quadrants, the 
Local Arm, and the Carina Arm are selected and their average peculiar motions
are calculated and compared with those of the YOCs and O--B2-type stars, as listed in Tables~\ref{tab:UVW_arms} and~\ref{tab:UVW_arms_std}. Due to the limited number of samples, the average peculiar motions from Table~\ref{tab:UVW_arms} are not significant at a $3 \sigma$ level, but they still reflect the potential kinematic characteristics of young objects located in the spiral arms.

As shown in Figure~\ref{fig:OC_OB},
in the Perseus Arm, 18 YOCs and 40 O--B2-type stars are located in the second quadrant, and 13 YOCs and 11 O--B2-type stars are located in the third quadrant.
This distribution is consistent with the fact that there are fewer and generally faint clouds in the third quadrant of the Perseus Arm~\citep{Dame+etal+2001}.
75\% YOCs and 73\% O--B2-type stars have positive values for the $U_{s}$ component, and 81\% YOCs and 76\% O--B2-type stars have negative values for the $V_{s}$ component.
As listed in Table~\ref{tab:UVW_arms}, 
the YOCs and O--B2-type stars in the Perseus Arm have positive average values of the $U_{s}$ component and negative average values of the $V_{s}$ component, indicating that the young objects are likely to have a tendency to move towards the GC and move slower than Galactic rotation. It is in line with the previous result obtained from masers~\citep{Sakai+etal+2019,Zhang+etal+2019}.
Besides, on average, the O--B2-type stars may exhibit larger peculiar motions than the masers and YOCs. 
When separating the O- and B0--B2-type stars, we discover that the $U_{s}$ and $V_{s}$ components of the peculiar motions of the O-type stars may be smaller than those of the B0--B2-type stars, while the uncertainties of O-type stars are large.

In the Local arm, 64 YOCs and 44 O--B2-type stars are located in the second quadrant, and 76 YOCs and 8 O--B2-type stars in the third quadrant.
57\% YOCs and 75\% O--B2-type stars have positive values for the $U_{s}$ component, 53\% YOCs and 71\% O--B2-type stars have negative values for the $V_{s}$ component. Combining with the average peculiar motions of young objects (listed in Table~\ref{tab:UVW_arms}), it is likely that O--B2-type stars, on average, present a movement toward the GC and a lagging motion with respect to the Galactic rotation. YOCs may have the same kinematic features as O--B2-type stars, but they are not significant.
The sources in the Local Arm may have smaller motions toward the GC than those of sources in the Perseus Arm.
The average peculiar motions of the masers in the Local Arm may be larger than those of the YOCs, but smaller than those of the O--B2-type stars in the $V_{s}$ component.
The B0--B2-type stars seem to present larger peculiar motions than the O-type stars. 

In the Carina arm, 51\% YOCs and 61\% O--B2-type stars have negative values of the $U_{s}$ component, 51\% YOCs and 59\% O--B2-type stars have positive values of the $V_{s}$ component.
The O--B2-type stars in the Carina Arm may have different kinematic features, in that they appear to be moving towards the Galactic anticenter. The O--B2-type stars appear to have no lagging motion and may even be ahead of Galactic rotation. The average values for the $U_{s}$ and $V_{s}$ components are $0.3 \pm 2.0$~\kms~and $-2.3 \pm 1.9$~\kms. Thus, the YOCs appear to have no peculiar motion in the radial direction and co-rotate with the Galactic orbits.
The average $V_{s}$ component of the masers in the Carina Arm is close to that of the young stars, and they may not exhibit lagging motion in relative to Galactic rotation. 
Contrary to the features in the Perseus and Local Arms, the peculiar motions of the O-type stars in the Carina Arm may be larger than those of the B0--B2-type stars.

The average peculiar motions and std values of the O--B2-types may be larger than those of YOCs or masers in the spiral arms near the Sun.
Additionally, in each spiral arm, the B0--B2-type stars perhaps show larger average peculiar motions than the O-type stars. 
Interestingly, the peculiar motions of sources located in the Perseus and Local Arms may be larger than those in the Carina Arm, implying that the peculiar motions of the young objects at different Galactocentric radii may be different. This issue will be investigated in the next subsection.

\tabletypesize{\normalsize}
\begin{deluxetable}{ccccccc}[htbp]
\tablecaption{Average Peculiar Motions of the Sources in the Individual Spiral Arms
\label{tab:UVW_arms}
}
\tablehead{
    Arm & Component & YOCs & O--B2 & O & B0--B2 & Maser
    }
\startdata
    \multirow{4}[0]{*}{Perseus} & $\overline{U_{s}}$ (\kms) & $6.0 \pm 2.6$ & $8.8 \pm 3.1$ & $-3.8 \pm 6.7$ & $13.1 \pm 3.1$ & $8.9 \pm 2.4$ \\
    & $\overline{V_{s}}$ (\kms) & $-6.5 \pm 1.6$ & $-8.5 \pm 2.3$ & $-4.0 \pm 6.2$ & $-10.0 \pm 2.1$ & $-8.0 \pm 2.5$ \\
    & $\overline{W_{s}}$ (\kms) & $0.3 \pm 0.7$ & $-0.7 \pm 0.7$ & $-1.2 \pm 2.0$ & $-0.5 \pm 0.6$ & $0.7 \pm 1.7$\\
    & Num & 31 & 51 & 13 & 38 & 28 \\ \hline
    \multirow{4}[0]{*}{Local} & $\overline{U_{s}}$ (\kms) & $1.8 \pm 0.7$ & $6.5 \pm 1.5$ & $3.1 \pm 4.0$ & $7.3 \pm 1.5$ & $-0.4 \pm 1.8$\\
    & $\overline{V_{s}}$ (\kms) & $-1.8 \pm 0.7$ & $-12.2 \pm 2.7$ & $-6.8 \pm 6.9$ & $-13.3 \pm 2.9$ & $-7.9 \pm 1.0$ \\
    & $\overline{W_{s}}$ (\kms) & $0.3 \pm 0.3$ & $-0.7 \pm 0.7$ & $0.8 \pm 1.8$ & $-1.0 \pm 0.7$ & $1.8 \pm 1.6$ \\
    & Num & 140 & 52 & 9 & 43 & 28 \\ \hline
    \multirow{4}[0]{*}{Carina} & $\overline{U_{s}}$ (\kms) & $0.3 \pm 2.0$ & $-4.8 \pm 2.0$ & $-9.0 \pm 3.7$ & $-2.6 \pm 2.4$ & $3.7 \pm 1.7$\\
    & $\overline{V_{s}}$ (\kms) & $-2.3 \pm 1.9$ & $4.5 \pm 1.8$ & $-3.3 \pm 3.2$ & $8.5 \pm 1.9$ & $0.4 \pm 1.4$ \\
    & $\overline{W_{s}}$ (\kms) & $0.0 \pm 0.8$ & $-1.7 \pm 0.5$ & $-1.7 \pm 1.1$ & $-1.6 \pm 0.6$ & $-4.2 \pm 1.2$ \\
    & Num & 35 & 85 & 29 & 56 & 17 \\
\enddata
\end{deluxetable}

\tabletypesize{\normalsize}
\begin{deluxetable}{ccccccc}
\tablecaption{Std Values of the Peculiar Motions of Sources in the Individual Spiral Arms
\label{tab:UVW_arms_std}
}
\setlength{\tabcolsep}{4mm}
\tablehead{
    Arm & Component & YOCs & O--B2 & O & B0--B2 & Maser
    }
\startdata
    \multirow{3}[0]{*}{Perseus} & $\sigma_{U_{s}}$ (\kms) & 14.7 & 21.8 & 24.2 & 19.1 & 13.0 \\
    & $\sigma_{V_{s}}$ (\kms) & 8.8 & 16.3 & 22.5  & 13.2 & 13.2 \\
    & $\sigma_{W_{s}}$ (\kms) & 3.7 & 4.7 & 7.0 & 3.5 & 9.0 \\ \hline
    \multirow{3}[0]{*}{Local} & $\sigma_{U_{s}}$ (\kms) & 8.3 & 10.6 & 11.9 & 10.1 & 9.4 \\
    & $\sigma_{V_{s}}$ (\kms) & 8.7 & 19.5 & 20.6 & 19.1 & 5.5 \\
    & $\sigma_{W_{s}}$ (\kms) & 4.1 & 4.7 & 5.3 & 4.5 & 8.4 \\ \hline
    \multirow{3}[0]{*}{Carina} & $\sigma_{U_{s}}$ (\kms) & 11.7 & 18.8 & 19.9 & 17.9 & 6.9 \\
    & $\sigma_{V_{s}}$ (\kms) & 11.2 & 16.4 & 17.4 & 14.3 & 5.8 \\
    & $\sigma_{W_{s}}$ (\kms) & 4.9 & 4.7 & 5.7 & 4.1 & 5.1 \\
\enddata
\end{deluxetable}

\clearpage

\subsection{Peculiar Motions Variations along Galactocentric Radius}

\begin{figure}[htbp]
    \centering
    \includegraphics[width = 5.5cm]{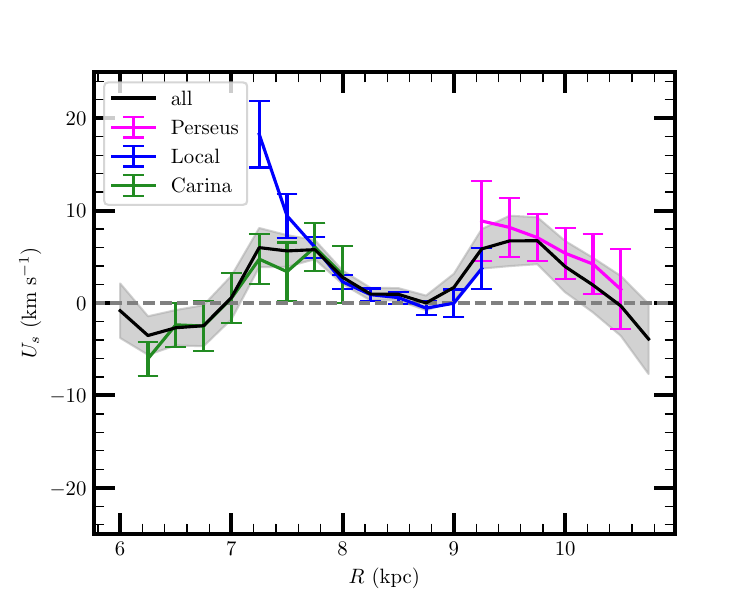}
    \includegraphics[width = 5.5cm]{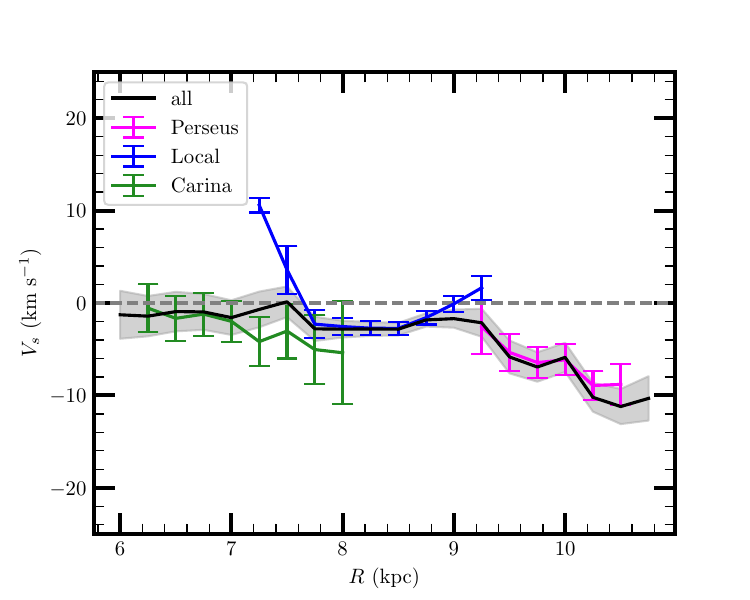}
    \includegraphics[width = 5.5cm]{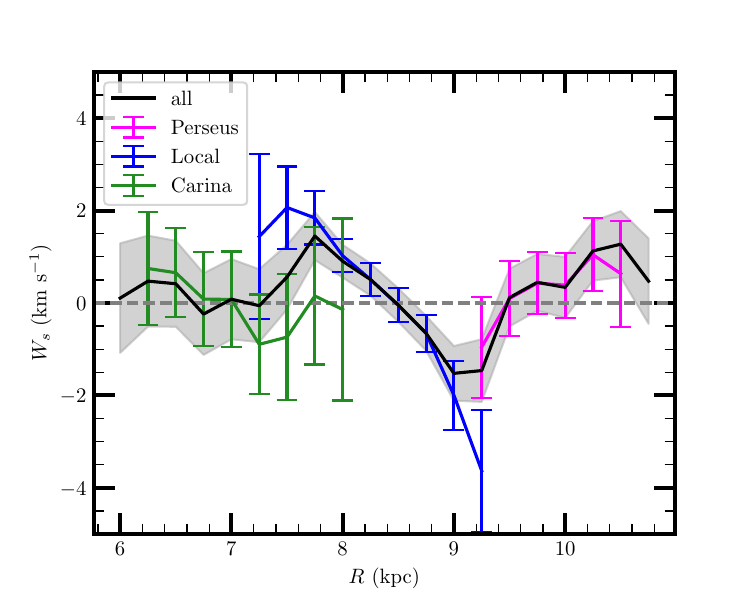}
    \caption{The average peculiar motions of the YOCs as a function of Galactocentric distance, $R$. The black solid lines indicate all sources, and shaded area represent the uncertainties. The magenta, blue, and green lines represent the sources located in the Perseus, Local, and Carina Arms. The average is calculated within a window of 1~kpc width in steps of 0.25~kpc, and $R_{0}$ is 8.15~kpc. Each interval must have more than five sources.}
    \label{fig:UVW_R_OC}
\end{figure}

\begin{figure}[htbp]
    \centering
    \includegraphics[width = 5.5cm]{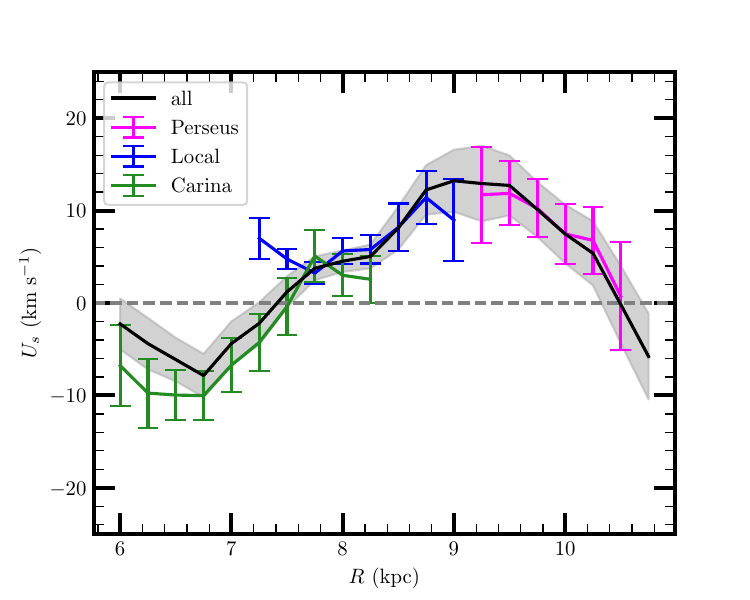}
    \includegraphics[width = 5.5cm]{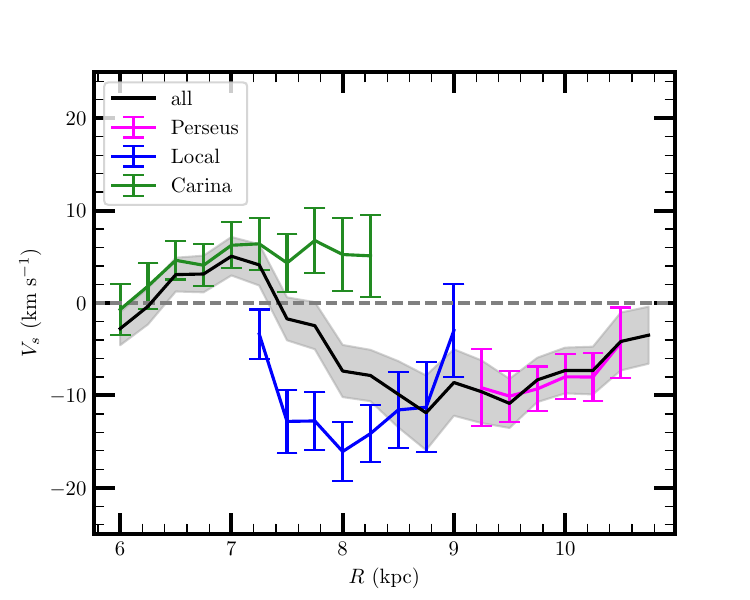}
    \includegraphics[width = 5.5cm]{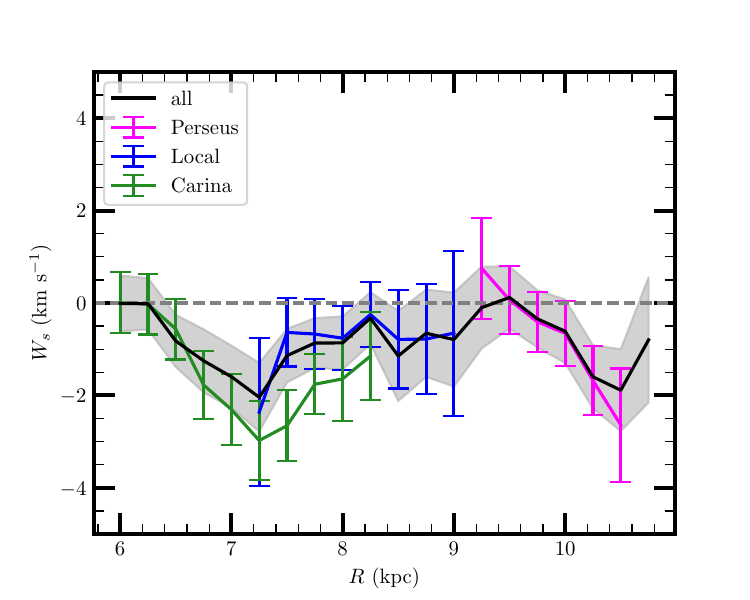}
    \caption{The average peculiar motions of the O--B2-type stars as a function of  Galactocentric distance, $R$. The others are the same as Figure~\ref{fig:UVW_R_OC}.}
    \label{fig:UVW_R_OB}
\end{figure}

To study the peculiar motions of the YOCs and O--B2-type stars with respect to Galactocentric radius, we analyze the average peculiar motion as a function of 
the Galactocentric radius in the range 6--11~kpc, as presented in 
Figures~\ref{fig:UVW_R_OC}--\ref{fig:UVW_R_OB}.
Since sources located in the two spiral arms may have the same Galactocentric distance, we show the average peculiar motions variations along the Galactocentric radius for all sources and sources located in different spiral arms, respectively.
Here, the average peculiar motion of each type of sources is calculated within a window of 1~kpc width in steps of 0.25~kpc. 

In these figures, the undulations of the 
peculiar motions of the O--B2-type stars as a function of Galactocentric radius are stronger than those displayed by the YOCs.
The $U_{s}$ component of YOCs may have two ``peak", which are located at $R \sim 7.0$~kpc and $R \sim 9.0$~kpc. However, considering the uncertainties of the values, the peaks are not significant. Thus, more YOCs are needed to further verify them.
The $U_{s}$ component of the O--B2-type stars increases from $R \sim$~7.0~kpc and reaches 
a maximum of $\sim$~14~\kms~at a distance of $R \sim 9.0$~kpc. 
For the $V_{s}$ components of the YOCs, they show a flat distribution from 6.0 to 9.0~kpc, and decrease beyond 9.0~kpc. For the $V_{s}$ components of the O--B2-type stars, they decline from 7.0 to 8.5~kpc and reach minimum of $\sim -12$~\kms~at a distance of $R \sim 8.5$~kpc. 
Outside 7.5~kpc, the $V_{s}$ component of the YOCs and O--B2-type stars remain smaller than zero, indicating the young objects may rotate slower than Galactic rotation.
The YOCs are more massive than single O-B2-type stars and would therefore better represent the average motion. Their declining $V_{s}$ velocity as a function of radius beyond 9.0~kpc could also suggest that the assumed rotation curve may overestimete the rotation velocity of the Milky Way.
As for the $W_{s}$ components of the YOCs and O--B2-type stars, their undulations are both weak and do not exceed $\pm~2$~\kms~in the Galactocentric range of 6--11~kpc.

In the Perseus Arm, the $U_{s}$ components appear to decrease with the Galactocentric radius for both YOCs and O--B2-type stars. The $V_{s}$ components of YOCs appear to decrease with Galactocentric radius, but increase with the radius for the O--B2-type stars. In the Local Arm, both the $U_{s}$ and $V_{s}$ components appear to show a ``U" structure along the Galactocentric radius for YOCs and O--B2-type stars. The lowest point of ``U" structure is at about 8.0~kpc. In the Carina Arm, the $U_{s}$ components of YOCs and O--B2-type stars appear to increase as the Galactocentric radius and would pass through the zero point at 7.0--7.5~kpc. The $V_{s}$ components of YOCs and O--B2-type stars do not appear to fluctuate significantly with the Galactocentric radius. 

Between 7.0 and 8.0~kpc, the sources located in the Carina Arm and the Local Arm may have the same Galactocentric radius. We further investigate the kinematic features of sources located in different spiral arms while having the same Galactocentric radius. The $U_{s}$ components decrease in this range for sources located in the Local Arm, while they increase for sources located in the Carina Arm. They would intersect at around 7.5~kpc. The $V_{s}$ components of the YOCs located in the Local Arm are larger than those in the Carina Arm. O--B2-type stars seem to exhibit the opposite phenomenon of YOCs. Both young objects indicate that the kinematic features may be different for the sources located in the two arms, while it needs to be confirmed with more young stars.

\clearpage
\section{Kinematic Distance}

Traditionally, one can construct a simple model of the rotation speeds of sources as a function of distance from the center of the Milky Way. Then, if one measures the line-of-sight velocity (radial velocity) of an object, one can determine its distance by matching the observation with the model prediction, which yields a kinematic distance.
However, the distances of sources derived from kinematic methods may be unreliable due to noncircular (peculiar) motions.
The significant peculiar motions of the sources are expected to affect their estimated kinematic distance.
\citet{Xu+etal+2006} found a parallax distance of $1.95 \pm 0.04$~kpc of the high-mass star-forming region W3(OH). The kinematic distance (i.e., $>$ 4~kpc) was more than twice as far as the parallax distance.
Hence, it is crucial to correct the kinematic model in regions where the peculiar motions
of sources are prominent to obtain relatively precise kinematic distance estimations.

First, we convert the heliocentric radial velocities of the YOCs and O--B2-type stars 
to the local standard of rest (LSR) velocities. 
The solar motion values derived from stars in different evolutionary stages are
different~\citep{Binney+Tremaine+2008}.
Since the objects discussed here are mainly young stars, we do not use the IAU solar motion. 
Instead, we use a ``revised LSR velocity" ($V_{\rm LSR}$) by applying the updated solar motion values obtained from masers~\citep{Reid+etal+2019} to estimate the kinematic distance of each object, which can be calculated by
\begin{equation}
    V_{\rm LSR} = V_{\rm Helio} + (V_{\odot} \sin l + U_{\odot} \cos l) \cos b + W_{\odot} \sin b,
    \label{eq:vh2vlsr}
\end{equation}
where $V_{\rm Helio}$ is the heliocentric radial velocity, $l$ and $b$ are the Galactic 
longitude and latitude, and \Usun, \Vsun, and \Wsun~are the solar motions. For a source with a given Galactic longitude and distance from the Sun, we can calculate its circular motion, $\Theta$, in the Galactic plane using the universal rotation curve~\citep{Persic+etal+1996,Reid+etal+2019}. 

\begin{figure}[htbp]
    \centering
    \includegraphics[width = 5.5cm]{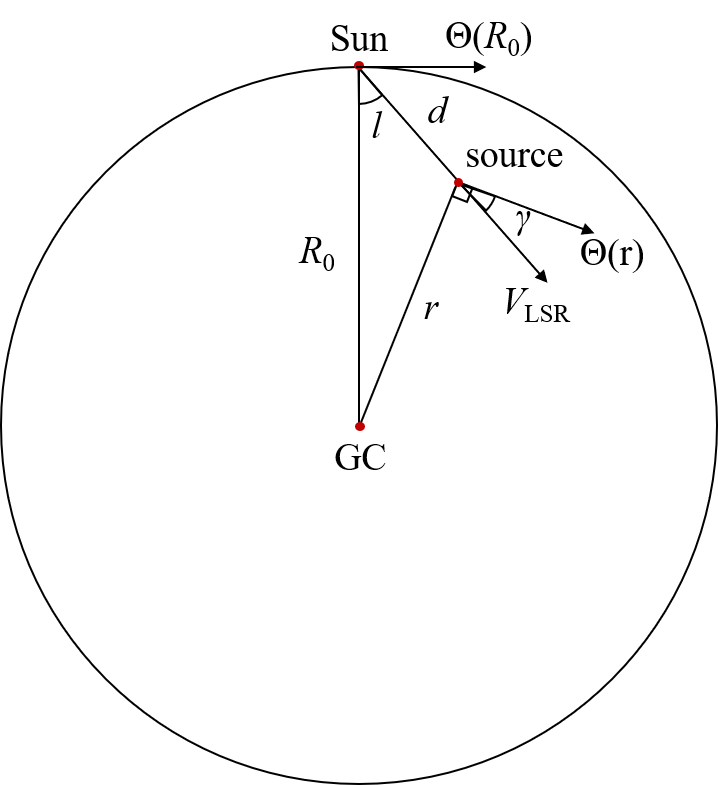}
    \caption{Schematic depiction of the source and Galactic parameters.}
    \label{fig:frame}
\end{figure}

The radial velocity of a source is the key to determining its kinematic distance.
Here, we define an assumed LSR radial velocity in a pure Galactic circular orbit, $V_{\rm LSR}^{\rm mod}$, which can be calculated by
\begin{equation}
    V_{\rm LSR}^{\rm mod} = (\Theta (r) \cos \gamma - \Theta (R_{0}) \sin l) * \cos b,
    \label{eq:mod}
\end{equation}
where $\Theta (r)$ is the theoretical rotation velocity at Galactocentric radius $r$ calculated by using the rotation curve. $\gamma$ is the angle between the line-of-sight direction and the rotation direction of the source, and $R_{0} = 8.15$~kpc~\citep{Reid+etal+2019}. 
A schematic depiction of these parameters is given in Figure~\ref{fig:frame}.
In contrast, if we know the LSR velocity of a source, its kinematic distance, $d$, can be derived by Equation~(\ref{eq:mod}). 

We adopt the probability density function (PDF, $P$) to estimate the kinematic distance. Following \citet{Reid+etal+2016}, we calculate it via
\begin{equation}
    P = \frac{1 - \exp[-(\Delta V_{\rm LSR} / \sigma_{V_{\rm LSR}})^{2} / 2]}{(\Delta V_{\rm LSR} / \sigma_{V_{\rm LSR}})^{2} / 2},
    \label{eq:pdf}
\end{equation}
where $\Delta V_{\rm LSR}$ is the difference between the LSR velocity and that expected from the rotation model of the Galaxy, $\sigma_{V_{\rm LSR}}$ is the uncertainty of the LSR velocity. Uncertainties in the kinematic distances include the 1$\sigma$ width of the PDF and uncertainties caused by the parameters in the rotation curve, i.e., a2 ($0.96 \pm 0.05$) and a3 \citep[$1.62 \pm 0.02$,][]{Reid+etal+2019}.

\begin{figure}[htbp]
    \centering
    \subfigure[]{\includegraphics[width = 8cm]{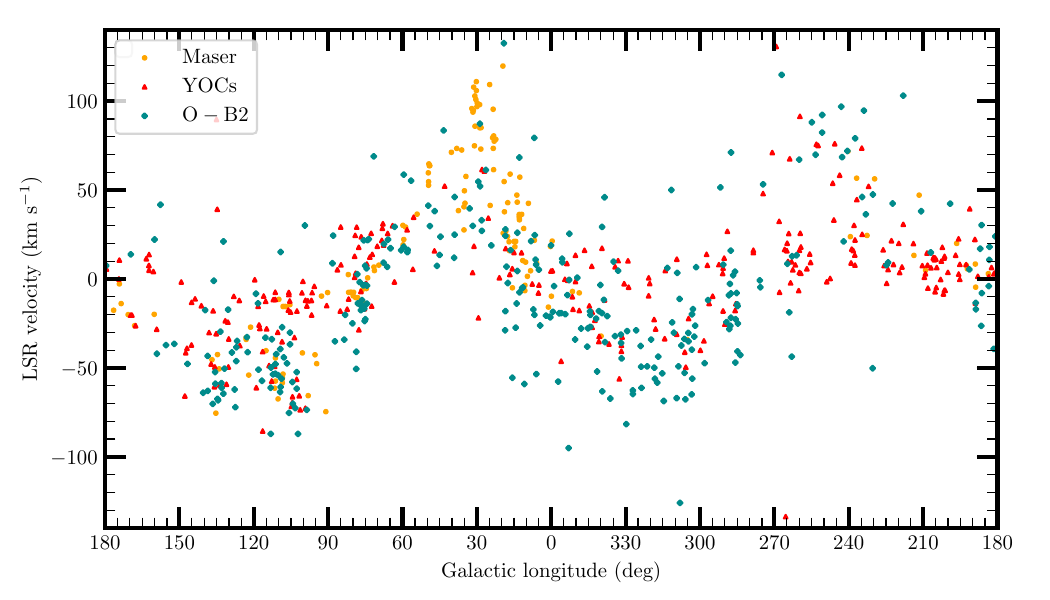}}
    \subfigure[]{\includegraphics[width = 8cm]{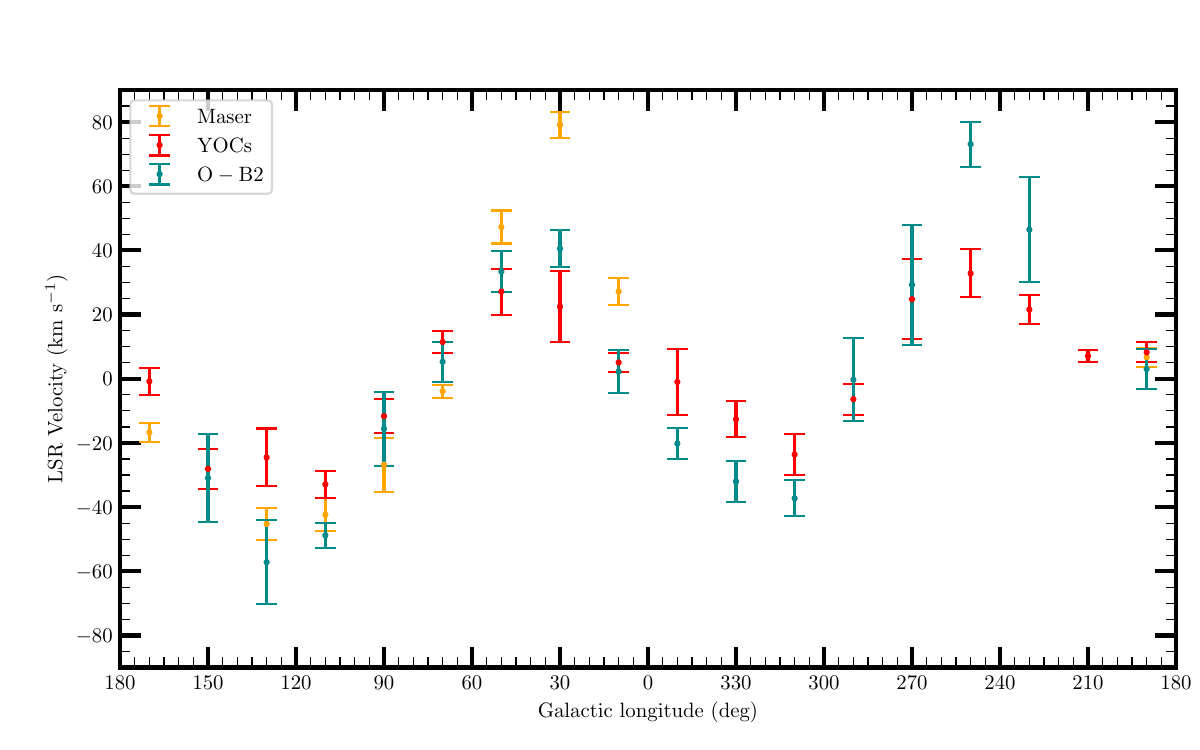}}
    \caption{(a): $l$--$v$ diagram. Colored dots are the sources of masers (orange dots), YOCs (red triangles), and O--B2-type stars (cyan diamonds). (b): variations in the average $V_{\rm LSR}$ per 20$^{\circ}$ for the masers (orange), YOCs (red), O--B2-type stars (cyan) along Galactic longitude. Each interval must have more than five sources.}
    \label{fig:lv_all}
\end{figure}

Figure~\ref{fig:lv_all}~(a) presents the distributions of YOCs, O--B2-type stars, and masers in a longitude--velocity ($l$--$v$) diagram.
Since YOCs and O--B2-type stars are mainly located within 6~kpc from the Sun, only masers in this range are selected for comparison.
The distributions of the YOCs and O--B2-type stars are roughly consistent with that of the  masers in the $l$--$v$ diagram, although they all have large dispersions. 
We calculate the average LSR radial velocities of the masers, YOCs, and O--B2-type stars with a Galactic longitude step every 20$^{\circ}$, as shown Figure~\ref{fig:lv_all}~(b).
Overall, the LSR velocities of the three young populations are concordant in most locations.
Near the Galactic longitude of $30^{\circ}$, the large differences in the LSR velocities among the three groups may be due to a lack of YOCs and O--B2-type stars (see Figure~\ref{fig:lv_all} (a)).
Therefore, we can combine the YOCs and O--B2-type stars to revise the model of kinematic distance estimation.

\begin{figure}[htbp]
    \centering
    \includegraphics[width = 12cm]{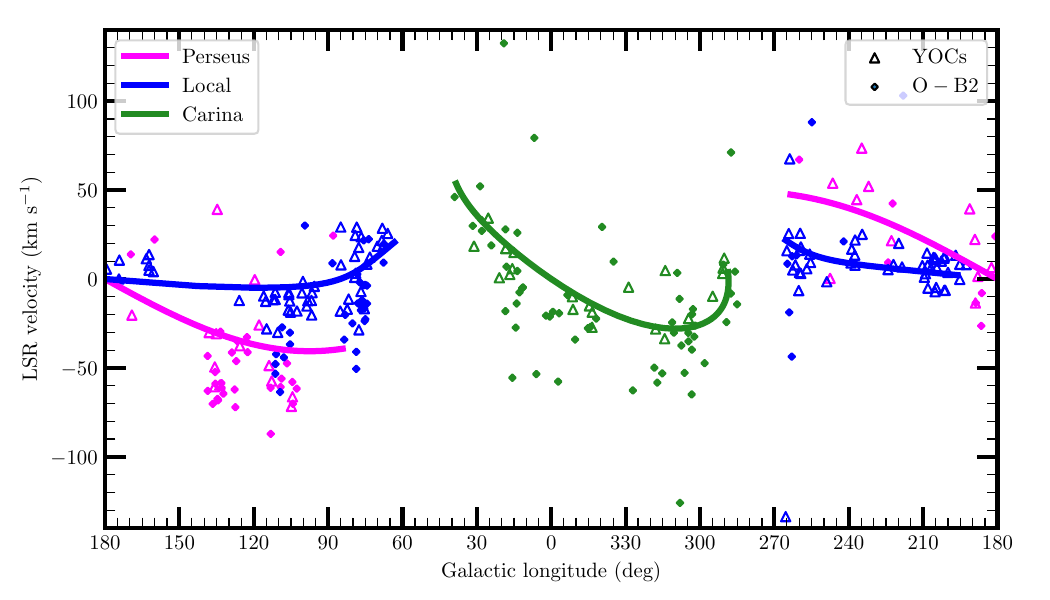}
    \caption{$V_{\rm LSR}$ values of the YOCs (triangles) and O--B2-type stars (diamonds) being assigned to the corresponding spiral arms (lines) as a function of Galactic longitude. The magenta, blue, and green lines represent the Perseus Arm, the Local Arm, and the Carina Arm, respectively, whose $V_{\rm LSR}^{\rm mod}$ values are calculated by using Equation~(\ref{eq:mod}).}
    \label{fig:lv_arms}
\end{figure}

In Figure~\ref{fig:lv_arms}, we display the distributions of the three spiral arms near the Sun in the $l$--$v$ diagram and of the YOCs and O--B2-type stars located within $1 \sigma$ from the centers of the spiral arms in the Galactic plane for comparison.
Here, the spiral arms are from the model presented by~\citet{Xu+etal+2023}, and 
$V_{\rm LSR}^{\rm mod}$ of each spiral arm is calculated using Equation~(\ref{eq:mod}).
In the $l$--$v$ diagram, the selected YOCs and O--B2-type stars are expected located 
near the spiral arms, while there are significant and systematic discrepancies between their measured $V_{\rm LSR}$ (using Equation~(\ref{eq:vh2vlsr}), shown as the colored points in Figure~\ref{fig:lv_arms}) and $V_{\rm LSR}^{\rm mod}$ values (the colored lines in Figure~\ref{fig:lv_arms}). 
It is caused by the peculiar motion of sources.
This suggests that the kinematic distances of these objects directly obtained from radial velocities deviate from their parallax distances. 
Therefore, in order to obtain more reliable kinematic distances for these sources, one can consider deducting the peculiar motions by adding offsets to radial velocities during the estimation.

Considering that the amount of YOCs and O--B2-type stars is limited
and that they are mainly concentrated in spiral arms, we focus on correcting the kinematic distance model toward the regions where the LSR velocity discrepancies between the sources and spiral arms are significant.
For example, near the direction of $l \sim 60^{\circ} - 90^{\circ}$ in the Local Arm, sources with LSR velocities less than $-10$~\kms~have significant discrepancies between $V_{\rm LSR}$ and $V_{\rm LSR}^{\rm mod}$, as shown in Figure~\ref{fig:lv_arms}.
After an offset of approximately 28~\kms~is applied to those sources, their distributions are close to the Local Arm in the $l$--$v$ diagram. Hence, this offset can yield more reliable kinematic distance estimations for these sources.
In total, six regions are found to need compensation via LSR velocity offsets for accurate kinematic distance estimations (see the expected values of these offsets in~Table~\ref{tab:rev}). 
Although the offsets are determined using sources in the spiral arms, in practice, it is not necessary to consider whether the source is located in a spiral arm, but only the conditions listed in Table~\ref{tab:rev}, i.e., Galactic longitude and radial velocity.

\begin{figure}
    \centering
    \subfigure[O--B2-type stars and YOCs]{\includegraphics[width = 8cm]{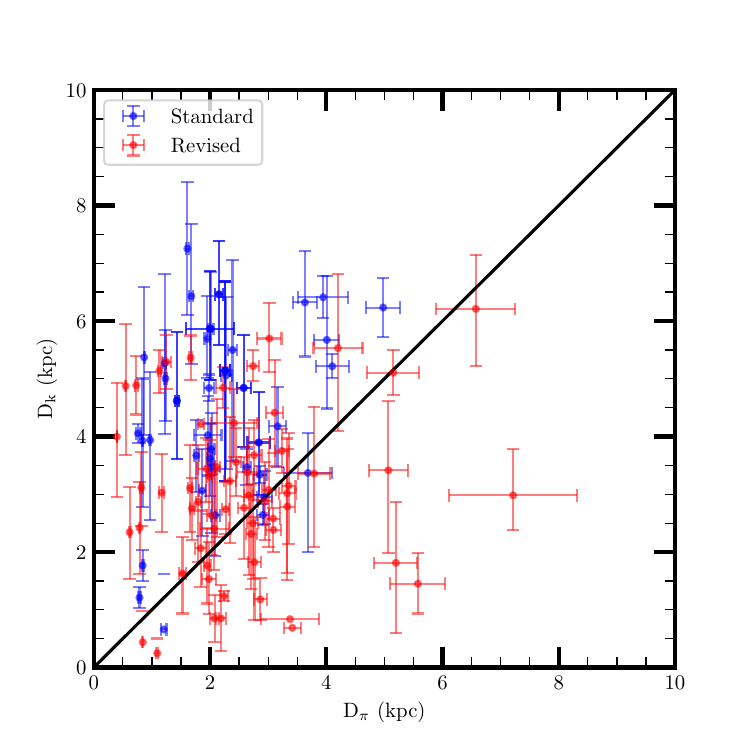}}
    \subfigure[Masers]{\includegraphics[width = 8cm]{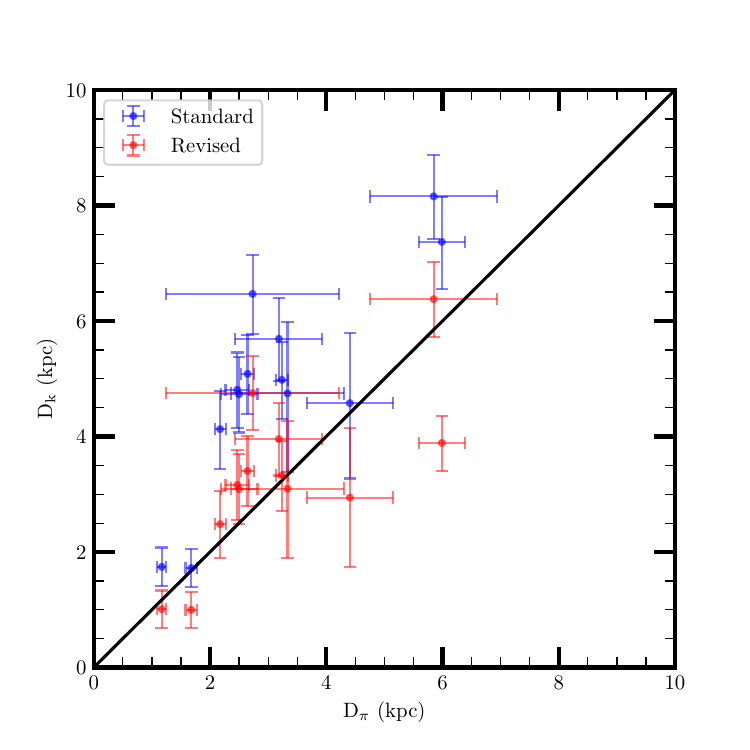}}
    \caption{Comparison between the parallax and kinematic distances. The young stellar population is shown in the left panel, and the masers are shown in the right panel. $D_{\pi}$ is the distance converted from the measured parallax. $D_{k}$ is the kinematic distance. The standard kinematic distance is obtained without considering the peculiar motion. The revised kinematic distance is obtained by adding the offsets to radial velocity to counteract the peculiar motion.}
    \label{fig:dist}
\end{figure}

The offsets are derived from the YOCs and O--B2-type stars located within 1$\sigma$ from the center of the spiral arms in the Galactic plane. Then we apply them to all young objects that meet the criteria in Table~\ref{tab:rev}. The parallax distances, standard distances, and revised distances of the YOCs and O--B2-type stars are presented in Figure~\ref{fig:dist}(a). Here, the standard kinematic distances are estimated without considering peculiar motion, while the revised kinematic distances are obtained by adding the offsets listed in~Table~\ref{tab:rev} to counteract the peculiar motion. For the YOCs and O--B2-type stars, their standard kinematic distances are generally greater than their parallax distances, while their revised kinematic distances are closer to their parallax distances. Moreover, we divide the YOCs and O--B2-type stars randomly into two groups, and calculate the standard kinematic distances for one group and the revised kinematic distances for the other group. After 100 re-samples and repetitions of the experiment, the average median value of the offset between standard kinematic distance and the parallax distance is 2.2~kpc, and average dispersion is 1.6~kpc. The corresponding values for the revised kinematic distance and parallax distance are 0.2 and 1.9~kpc, respectively. The typical relative uncertainty of the kinematic distance for these young objects is about 0.3. Furthermore, for the offsets in correction that are listed in~Table~\ref{tab:rev}, we check them using the masers, since their distribution are consistent with those of the YOCs and O--B2-type stars in the $l$--$v$ diagram (Figure~\ref{fig:lv_all}). As shown in~Figure~\ref{fig:dist}(b), compared with the standard kinematic distances, the revised kinematic distances of the masers are in agreement with their parallax distances, and the median offset is reduced from 2.0 to 0.3~kpc. The offsets, which are obtained from a fraction of YOCs and O--B2-type stars, are successfully applied not only all YOCs and O--B2-type stars, but also masers. It suggests that the method adopted here is effective.

The above results indicates that, within the range of 6~kpc from the Sun, if a source satisfies the conditions listed in Table~\ref{tab:rev}, one can add the corresponding offset to its observed radial velocity to determine a more reliable kinematic distance than the standard kinematic distance.
However, using the present-day data, we can make kinematic distance correction in only six regions in the solar neighborhood. With an increase of available data in the future, we are expected to modify the kinematic distance model in all regions near the Sun, and even in the distant regions of the Milky Way.

\section{Summary}
We investigate the kinematics of YOCs and O--B2-type stars in the local spiral structure. The YOCs and O--B2-type stars in the different spiral arms show different kinematic features. The young populations located in the outer spiral arms may exhibit larger peculiar motions than those in the inner spiral arms.
In the Perseus and Local Arms, the YOCs and O--B2-type stars perhaps move towards the GC and rotate slower than the rotation of the Milky Way. 
In contrast, the young populations in the Carina Arm may move toward the Galactic anticenter and rotate slightly faster of Galactic rotation.
Using the astrometric parameters of the young stellar populations, we also perform kinematic distance corrections for several regions in the solar neighborhood. At present, the corrections to the kinematic distance estimates are only valid up to 6 kpc from the Sun, yet it provides a useful reference for future kinematic distance estimates.

\begin{deluxetable}{ccc}[htbp]
\tablecaption{Offsets Used for the Kinematic Distance Estimation
\label{tab:rev}
}
\setlength{\tabcolsep}{10mm}
\tablehead{
    $l~(^{\circ})$ & condition & offset (\kms)
    }
\startdata
    60--90   & $V_{\rm LSR} < -10$~\kms & $~~~28$ \\
    90--130  & $V_{\rm LSR} < -45$~\kms & $~~~20$ \\
    130--150 & $V_{\rm LSR} < -50$~\kms & $~~~27$ \\
    230--250 & $V_{\rm LSR} > ~~50$~\kms & $-10$ \\
    250--280 & $V_{\rm LSR} < ~~~~0$~\kms & $~~~30$ \\ 
    290--330 & $V_{\rm LSR} < -40$~\kms & $~~~25$ \\
\enddata
\end{deluxetable}

\begin{acknowledgements}
    We appreciate the anonymous referee for the instructive comments that helped us to improve the paper. This work was funded by the NSFC Grands 11933011, National SKA Program of China (Grant No. 2022SKA0120103) and the Key Laboratory for Radio Astronomy. L.Y.J. thanks the support of the NSFC grant No. 12203104, the Natural Science Foundation of Jiangsu Province (grant No. BK20210999). This work has made use of data from the European Space Agency (ESA) mission \emph{Gaia} (\url{https://www.cosmos.esa.int/gaia}) processed by the \emph{Gaia} Data Processing and Analysis Consortium (DPAC, \url{https://www.cosmos.esa.int/web/gaia/dpac/consortium}). Funding for the DPAC has been provided by national institutions, in particular the institutions participating in the \emph{Gaia} Multilateral Agreement.
\end{acknowledgements}
        
\clearpage

\end{document}